\documentclass[usenatbib]{mnras}

\DeclareRobustCommand{\VAN}[3]{#2}
\let\VANthebibliography\thebibliography
\def\thebibliography{\DeclareRobustCommand{\VAN}[3]{##3}\VANthebibliography}

\usepackage{graphicx}	
\usepackage{amsmath}	
\usepackage{amssymb}	
\usepackage{newtxtext,newtxmath}
\usepackage{physics}
\usepackage{xspace}
\usepackage{psfrag}
\usepackage{slashbox}
\usepackage{epsfig}
\usepackage{pdfpages}
\usepackage[T1]{fontenc}

\newcommand{\be}{\begin{equation}}
\newcommand{\e}{\end{equation}}
\newcommand{\bear}{\begin{eqnarray}}
\newcommand{\ear}{\end{eqnarray}}

\newcommand{\del}{\partial}

\newcommand{\comment}[1]{}

\def\xi{x^{i}_{\ion{H}{i}}\,}
\def\xh1{x_{\ion{H}{i}\,}}
\def\xb{\bar{x}_{\ion{H}{i}}}
\def\xb2{\bar{x}_{\ion{H}{ii}}}

\def\Ph1{P_{\ion{H}{i}}}
\def\Bh1{B_{\ion{H}{i}}}
\def\eh1{\eta_{\ion{H}{i}}}

\def\aap{AAP}
\def\apj{ApJ}

\def\apjs{ApJS}
\def\apjl{ApJL}
\def\mnras{MNRAS}
\def\prd{PRD}

\def\P{\hat{P}}

\def\mp{\, {\rm Mpc}^{-1}}

\newcommand{\TB}{\delta T_{\rm b}}

\newcommand{\Omegam}{\Omega_{\rm m}}

\usepackage[normalem]{ulem}

\title[CD 21-cm bispectrum]{Probing IGM Physics during Cosmic Dawn using the Redshifted 21-cm Bispectrum}

\author[Kamran et al.]{Mohd Kamran$^{1}$\thanks{E-mail: kamranmohd080@gmail.com},
Suman Majumdar$^{2,3}$,
Raghunath Ghara$^{4}$,
Garrelt Mellema$^{5}$, \newauthor
Somnath Bharadwaj$^{6}$,
Jonathan R. Pritchard$^{3}$,
Rajesh Mondal$^{7}$,
Ilian T. Iliev$^{8}$
\\
$^{1}$Theoretical astrophysics, Department of Physics and Astronomy, Uppsala University, Box 516, 751 20 Uppsala, Sweden\\
$^{2}$Department of Astronomy, Astrophysics \& Space Engineering, Indian Institute of Technology Indore, Indore 453552, India\\
$^{3}$Department of Physics, Blackett Laboratory, Imperial College, London SW7 2AZ, U. K.\\
$^{4}$ARCO (Astrophysics Research Center), Department of Natural Sciences, The Open University of Israel, 1 University Road, PO Box 808, Ra'anana 4353701, Israel\\
$^{5}$The Oskar Klein Centre and The Department of Astronomy, Stockholm University, AlbaNova, SE-10691 Stockholm, Sweden\\
$^{6}$Department of Physics, Indian Institute of Technology Kharagpur, Kharagpur - 721302, India\\
$^{7}$School of Physics and Astronomy, Tel Aviv University, Tel Aviv, 69978, Israel.\\
$^{8}$Department of Physics and Astronomy, Pevensey II Building, University of Sussex, Brighton BN1 9QH, UK\\
}

\date{Accepted XXX. Received YYY; in original form ZZZ}

\begin{document}
\label{firstpage}
\pagerange{\pageref{firstpage}--\pageref{lastpage}}
\maketitle


\begin{abstract}
With the advent of the first luminous sources at Cosmic Dawn (CD), the redshifted 21-cm signal, from the neutral hydrogen in the Inter-Galactic Medium (IGM), is predicted to undergo a transition from absorption to emission against the CMB. Using simulations, we show that the redshift evolution of the sign and the magnitude of the 21-cm bispectrum can disentangle the contributions from Ly$\alpha$ coupling and X-ray heating of the IGM, the two most dominant processes which drive this transition. This opens a new avenue to probe the first luminous sources and the IGM physics at CD.    
\end{abstract}


\section{Introduction}
\label{sec:intro}

The Cosmic Dawn (CD) and Epoch of Reionization (EoR) are the periods in the history of the Universe during which the formation of the first sources of radiation caused major changes in the thermal and ionization states of the Inter-Galactic Medium (IGM). Various complex astrophysical processes dictate the state of the IGM during CD \citep{pritchard08,pritchard12}. 

The redshifted 21-cm line, originating from the hyperfine spin-flip transition of the electron in the ground state of neutral hydrogen atoms (HI),   traces the cosmological and astrophysical evolution of the IGM during CD.
Ongoing and upcoming radio interferometric experiments e.g.
GMRT \citep{paciga13}, 
LOFAR \citep{mertens20},
MWA \citep{barry19}, 
PAPER \citep{kolopanis19}, 
HERA \citep{deboer17} and 
SKA \citep{koopmans15} are expected to be sensitive enough to probe this signal 
through various statistics such as the variance \citep{iliev08, watkinson15}, power spectrum \citep{jensen13}, etc. 

The power spectrum measures the amplitude of fluctuations in a signal at different length scales and fully quantifies the statistical properties of a pure Gaussian random field. However, the CD $21$-cm signal is highly non-Gaussian due to the joint effects of the non-random distribution of first light sources in the IGM, the nature of the radiation emitted by these sources, as well as the interaction of this radiation with the HI gas in the IGM \citep{bharadwaj05a, mellema06, mondal15}. The nature and strength of fluctuations of this signal at any location in the IGM are determined by the spin temperature $T_{\rm S}$ defined through $n_1/n_0 = 3\exp(-0.068{\, \rm K}/T_{\rm S})$ where $n_1$ and $n_0$ are respectively the population densities of the excited and the ground states of the HI 21 cm transition. Different IGM processes, determined by the nature of the radiation from the first sources, affect $T_{\rm S}$ \citep{pritchard08}. The fluctuations in the HI 21-cm signal, and their  inherent non-Gaussianity, are expected to carry the signatures of these IGM processes  \citep{bharadwaj05a, majumdar18, watkinson19, hutter19, majumdar20, kamran21}. The power spectrum cannot fully describe this non-Gaussianity and does not capture  the complete IGM physics. It is necessary to consider higher-order statistics such as the bispectrum to capture the non-Gaussianity of the 21-cm signal \citep{majumdar18, giri19, watkinson19, hutter19, majumdar20, kamran21, ma21}.

Compared to the power spectrum, which is always positive by definition, the CD 21-cm bispectrum can be either positive or negative \citep{majumdar18, watkinson19, hutter19, majumdar20, kamran21}. In this letter, we demonstrate, for the first time, that it is possible to identify the dominant IGM processes through different stages of CD by studying the evolution of the sign and magnitude of the 21-cm bispectrum.

\section{Simulating the HI 21-cm signal from the Cosmic Dawn}
\label{sec:sim}
 This work is based on an N-body simulation using the {\sc cubep$^3$m} code \citep{Harnois12} in a volume  $(500\,h^{-1})^3$ comoving Mpc$^3$ using $6912^3$ dark matter particles. The resolved halos have been identified with having at least 25 particles, resulting in minimum halo mass $\approx 10^9$~M$_\odot$. The dark matter density and velocity fields were interpolated on a $600^3$ grid. We then use these ingredient fields in the $1$D radiative transfer code {\sc GRIZZLY} \citep{ghara18,ghara15a} to simulate the $21$-cm differential brightness temperature ($\TB$) maps at $22$ redshift snapshots in the range $z = 9$ to $20$ following the equations in \cite{pritchard08}:
\begin{equation}
\begin{split}
    \delta T_{\rm b}(\textbf{r},z) = 27x_{\rm HI}(\textbf{r},z)\big(1+\delta_{\rm b}(\textbf{r}, z)\big)\Big(1-\frac{T_{\rm CMB}(z)}{T_{\rm S}(\textbf{r},z)}\Big)  \\ \times \Big( \frac{\Omega_{\rm b}h^2}{0.023}\Big) \Big(\frac{0.15}{\Omega_{\rm m}h^2} \frac{1+z}{10}\Big)^{1/2}\Big( \frac{\del v_r /\del r}{(1+z)H(z)}  \Big)  {\, \rm mK} 
    \label{eq:tb}
\end{split}
\end{equation}
where $\textbf{r}$ is the comoving distance to the source of emission, $z$ is the redshift at which the signal was emitted, $\, x_{\rm HI}(\textbf{r},z)$ is the hydrogen neutral fraction, $\, \delta_{\rm b}(\textbf{r}, z)$ is the fluctuations in the underlying baryon density and $\, T_{\rm CMB} (z)$ is the Cosmic Microwave Background (CMB) radiation temperature and the final term reflects the peculiar velocity of the gas along the line of sight to the observer. The spin temperature is connected to the IGM processes via the equation
\begin{equation}
    {T_{\rm S}(\textbf{r},z)} = \frac{T_{\rm CMB}(z)+x_{\alpha}(\textbf{r},z)T_{\rm g}(\textbf{r},z)}{1+x_{\alpha}(\textbf{r},z)}
    \label{eq:ts}
\end{equation}
Hence the value of $T_{\rm S}$ will be determined by the strength of the Ly$\alpha$ coupling process ($x_\alpha$) and the temperature of the IGM ($T_{\rm g}$). The $T_{\rm g}$ is determined by the adiabatic cooling due to cosmological expansion and the heating due to X-ray sources. Here we do not consider the impact of collisions on $T_{\rm S}$ as our redshift range ($20 > z > 9$) is well below the regime where collisions are important ($z\gtrsim 30$) \citep{pritchard08}. Hence, the evolution of the fluctuations in the 21-cm signal depends on the changes in the Ly$\alpha$ coupling, X-ray heating and photo-ionization of the IGM gas and below we will focus on the impact of these three processes on the bispectrum. The 21-cm signal will be observed in either absorption or emission depending on whether the factor $(1-T_{\rm CMB}/T_{\rm S})$ is negative or positive, respectively. However, once heating has raised $T_{\rm g}$ substantially above $T_{\rm CMB}$ at the late stages of the CD, the 21-cm emission becomes insensitive to the value of $T_{\rm S}$, a regime known as spin temperature saturation is reached, and then photo-ionization becomes important.

Further, the X-ray sources in our simulations have a spectral energy density $I_X(E) \propto E^{-\alpha}$ with $\alpha = 1.5$ which roughly represents mini-QSOs \citep{gallerani17,martocchia17}. These are expected to be the major sources of soft X-ray photons during the CD. Note that the point regarding mini-QSOs being the major sources of X-rays during CD is at best controversial and debatable topic. We would like to emphasize that here we are not claiming that mini-QSOs are the major sources of X-ray photons, we have just assumed mini-QSOs as one of the possible sources of X-ray photons and it does not matter much what we assumed anyway \citep{kamran21}. Apart from the X-ray photons produced by mini-QSOs, we also consider the contributions from star-forming galaxies which produce ionizing photons. We further assume the stellar content of a galaxy to be proportional to the mass of the dark matter halo that hosts the galaxy. Thus, the stellar mass inside a halo of mass $M_{\rm halo}$ is  $M_\star=f_\star \left(\frac{\Omega_{\mathrm b}}{\Omegam}\right) M_{\rm halo}$ where  $f_\star$ is the star formation efficiency. We choose $f_\star = 0.03$ \citep{behroozi15, sun16} throughout this paper. More details about these simulations can be found in \cite{kamran21}. The cosmological parameters used in the simulations are $h = 0.7$, $\Omega_{\mathrm{m}} = 0.27$, $\Omega_{\Lambda} = 0.73$, $\Omega_{\mathrm{b}} = 0.044$, which are consistent with \textit{WMAP} \citep{hinshaw13} and \textit{Planck} results \citep{planck14}.


\begin{table}
\begin{tabular}{|l||*{4}{c|}}\hline
\backslashbox{Processes}{Scenarios}
&\makebox[2.7em]{Model-a$_{0}$}&\makebox[2.7em]{Model-a}&\makebox[2.6em]{Model-b}&\makebox[2.6em]{Model-c}\\\hline\hline
Ly$\alpha$-coupling &Yes&Yes&Saturated&Yes\\\hline
X-ray heating &No&No&Yes&Yes\\\hline
Ionization &No&Yes&Yes&Yes\\\hline
\end{tabular}
\caption{All simulated CD scenarios considered in this study.}
\label{table1}
\end{table}

\begin{figure*}
\begin{center}
\includegraphics[width=160mm,angle=0]{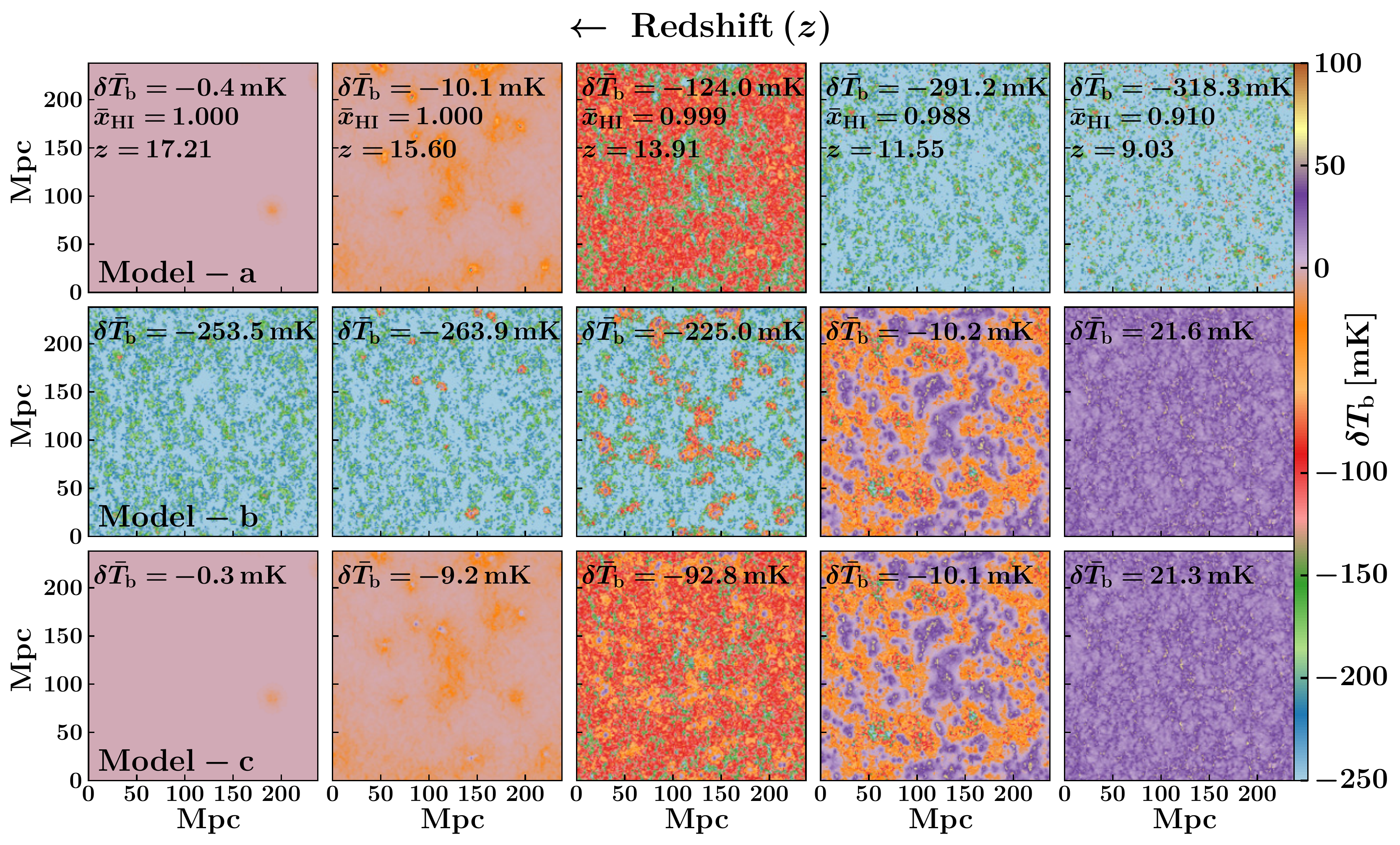}
\caption {Zoomed in slices (side length $238.09\, {\rm Mpc}$) of the brightness temperature showing the redshift evolution of the 21-cm signal for scenarios Model-a (Top panels), Model-b (Middle panels), Model-c (Bottom panels). The full simulation volume is $(714.29\, {\rm Mpc})^3$ in size.}
\label{fig:Tb_map}
\end{center}
\end{figure*}

{In this letter we consider four different scenarios for the evolution of the CD signal  as listed in Table \ref{table1}. The first three are simplistic and extreme scenarios in which only a single physical process dominates the fluctuations in the 21-cm signal. The fourth scenario includes all three physical processes and therefore is the most realistic CD scenario. We use the first three scenarios to demonstrate the unique signatures of each of the three physical processes on the 21-cm bispectrum, which eventually helps us in explaining the bispectrum from Model-c, the most realistic scenario.} The details of these scenarios are as follows: Model-a$_0$ incorporates only the evolution of Ly$\alpha$ coupling in a cold and neutral IGM. Model-a further includes the  photo-ionization of HI in the IGM. Model-b considers the processes of X-ray heating and ionization but assumes that a very strong Ly$\alpha$ background fully couples $T_{\rm S}$ to $T_{\rm g}$. Finally, Model-c includes all of these processes in a self-consistent manner. Figure \ref{fig:Tb_map} shows slices through the $\TB$ cube from very early to very late stages of the CD (left to right) for Model-a (top panels), Model-b (middle panels), and Model-c (bottom panels). The redshift evolution of the $\delta\bar{T}_\mathrm{b}$ (global signal) for all of these models are shown in the left panel of Figure \ref{fig:TbPSBS_z}.


\section{Bispectrum estimation}
\label{sec:bis_est}
We adopt the bispectrum estimator and associated algorithm discussed in \cite{majumdar18, majumdar20} to compute the bispectrum from the simulated data as
\begin{equation}
  \hat{B}_{i}({\bf k}_1, {\bf k}_2, {\bf k}_3) = \frac{1}{N_{{\rm tri}}V} \sum_{
[{\bf k}_1 +{\bf k}_2 +{\bf k}_3 = 0] \in i }\Delta\Tilde{T}_{\rm b}(\textbf{k}_1) \Delta\Tilde{T}_{\rm b}(\textbf{k}_2) \Delta\Tilde{T}_{\rm b}(\textbf{k}_3) \, ,
\label{eq:bispec_est}
\end{equation}
where $\Delta\Tilde{T}_{\rm b}(\textbf{k})$ is the Fourier transform of $\delta T_{\rm b}(\textbf{r})$, $V$ is the observation or simulation volume and $N_{\rm tri}$ is the number of closed triangles associated with the $i^{\rm th}$ triangle configuration bin while satisfying the condition ${\bf k}_1 +{\bf k}_2 +{\bf k}_3 =0$. Throughout this letter, we discuss spherically averaged bispectrum estimates obtained from our simulations.

\section{Results}
\label{sec:results}

\begin{figure*}
\begin{center}
\includegraphics[width=170mm,angle=0]{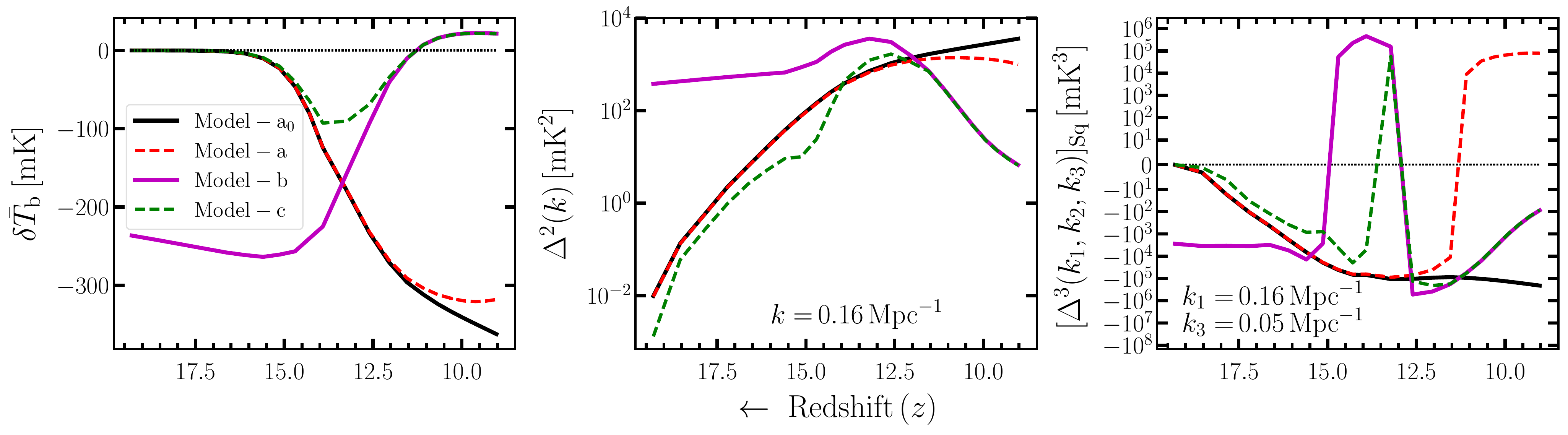}
\caption{Redshift evolution of the global 21-cm signal (left panel), the spherically averaged power spectrum at $k \sim 0.16 \mp$ (middle panel), and the bispectrum for squeezed limit $k$-triangle with $k_1 \sim 0.16 \mp$ (right panel). Different colours represent different CD scenarios as shown in Table \ref{table1}.}
\label{fig:TbPSBS_z}
\end{center}
\end{figure*}

We estimate the power spectrum [$P(k)$] and bispectrum [$B(k_1, k_2, k_3)$] for all CD scenarios of Table \ref{table1} for $k_1 = 0.16 \mp$, which represents large length scales that can be potentially probed by telescopes e.g. SKA. For the sake of convenience, we designate $k_1 \leq 0.16 \mp$ as `large-scale'. We normalize these statistics as $\Delta^{2}(k) = k_1^3 P(k)/2\pi^2$ and $\Delta^{3}(k_1, k_2, k_3) = k_1^3 k_2^3 B(k_1, k_2, k_3)/(2\pi^2)^2$. Recent studies \citep{watkinson21,mondal21} expect that among all $k$-triangles, the bispectrum for squeezed limit $k$-triangle to have the maximum signal-to-noise ratio and thus to have the highest probability of detection in future SKA observations. In addition to this, the squeezed limit bispectrum for $k_1 = k_2 = 0.16 \mp> k_3 \sim 0.05 \mp$, provides the correlation between the signal fluctuations at large and very large length scales. Motivated by this we only consider the large-scale squeezed limit bispectrum i.e. $[\Delta^3(k_1, k_2, k_3)]_{\rm Sq}$ and investigate how they are affected by various IGM processes. The right panel in Figure \ref{fig:TbPSBS_z} shows the redshift evolution of $[\Delta^3(k_1, k_2, k_3)]_{\rm Sq}$.


We first discuss the  expected signal using simple analytical models which help  to interpret the results from our simulations. We follow \cite{bharadwaj05a} and \cite{majumdar18} who modelled the EoR signal using a set of non-overlapping ionized bubbles, all  of the comoving radius $R$, embedded in a uniform, neutral, spin temperature saturated ($T_{\rm S} \gg T_{\rm CMB}$) IGM. The expected 21-cm signal  in the Fourier domain is of the form:   $\Delta\Tilde{T}_{\rm b} \propto -W(kR) \sum_n \exp{i\textbf{k}\cdot \textbf{r}_n}$, where $W(kR)$ is the spherical top-hat window function, the sum is over different spheres  and  $\textbf{r}_n$ their centers.
A large scales  $(W(kR) \approx 1)$, the signal is negative because each  ionized bubble appears as a decrement against a uniform background emission. Note that we only consider the sign here, so the exact shape of the regions is irrelevant.
Here we  have generalized this approach and applied it to different IGM processes which affect   the CD 21-cm signal at large scales. 

\begin{enumerate}
\item{A Ly$\alpha$ coupled region in a neutral, cold and approximately uncoupled ($T_\mathrm{S} \approx T_\mathrm{CMB}$) background. The region will have a strong negative 21-cm signal in a very weak negative background, whereby  $\Delta\Tilde{T}_{\rm b} $ is negative.}
\item{An X-ray heated region ($T_\mathrm{S}>T_\mathrm{CMB}$) in a neutral, cold and fully Ly$\alpha$ coupled background ($T_\mathrm{S}<T_\mathrm{CMB}$). The region will have a positive 21-cm signal in a negative background, whereby  $\Delta\Tilde{T}_{\rm b}$ is  positive.}
\item{An ionized region in a neutral, heated and fully Ly$\alpha$ coupled background ($T_\mathrm{S}>T_\mathrm{CMB}$). The region will have zero 21-cm signal in a positive background, whereby  $\Delta\Tilde{T}_{\rm b}$ is negative.}
\item{An ionized region in a neutral, cold and fully Ly$\alpha$ coupled background ($T_\mathrm{S}<T_\mathrm{CMB}$). The region will have zero 21-cm signal in a negative background, whereby  $\Delta\Tilde{T}_{\rm b}$ is  positive.}
\item{A Ly$\alpha$ coupled region in a neutral and heated background ($T_\mathrm{S}>T_\mathrm{CMB}$). The region will have a negative 21-cm signal in a positive background, whereby  $\Delta\Tilde{T}_{\rm b}$ is  negative.} 
\end{enumerate}
 
{As stated earlier, the large-scale squeezed limit bispectrum $[\Delta^3]_{\rm Sq}$ probes the correlation between three 
$\Delta\Tilde{T}_{\rm b}$. It is important to note that the sign of $[\Delta^3]_{\rm Sq}$ will reflect that of $\Delta\Tilde{T}_{\rm b}$.}


\subsection{Model-a\texorpdfstring{$_0$}{\_0} and Model-a}
\label{sec:modela0a}
The redshift evolution of the global signal $\delta\bar{T}_\mathrm{b}$, the spherically averaged power spectrum $\Delta^2$ and squeezed limit bispectrum $[\Delta^3]_{\rm Sq}$ for Model-a$_0$ are shown by solid black lines in the left, middle and right panels of Figure \ref{fig:TbPSBS_z}. This CD scenario only involves  Ly$\alpha$ coupling process. This process starts with a negligible contribution to the global 21-cm signal during very early stages and gradually grows the 21-cm fluctuations with decreasing redshift (left panel of Figure \ref{fig:TbPSBS_z}). The magnitude of the $\Delta\Tilde{T}_{\rm b}$ gradually enhances and consequently causes the magnitude of $[\Delta^3]_{\rm Sq}$ to increase with decreasing redshift. As expected from the analytical model 1, the sign of $[\Delta^3]_{\rm Sq}$ is indeed negative.

Considering Model-a (dashed red line in Figure \ref{fig:TbPSBS_z}), $[\Delta^3]_{\rm Sq}$ agrees with Model-a$_0$ for $z \ge 12$ where $\bar{x}_{\rm HI}  \ge 0.99$ and  the ionization process, included in Model-a but not in Model-a$_0$, are unimportant. Ionization gains importance at $z \lesssim 12$ where it causes a sign reversal in $[\Delta^3]_{\rm Sq}$ from negative to positive, consistent with analytical model 4.

\subsection{Model-b and Model-c}
\label{sec:modelbc}
Model-b (magenta lines in Figure \ref{fig:TbPSBS_z}) considers X-ray heating in an IGM with fully saturated Ly$\alpha$ coupling ($T_{\rm S} = T_{\rm g}$) where $\delta\bar{T}_\mathrm{b}$ transitions from  absorption to  emission. During the early stages ($19 \gtrsim z \gtrsim 15.5$), only a few X-ray sources are active, and  their  heating  cannot compete with the adiabatic cooling due to the expansion of the Universe. The net effect is a gradual cooling of the IGM which causes  $T_{\rm S}$ to decrease faster than $T_{\rm CMB} (\propto (1+z)$). This implies that $\big[1-T_{\rm CMB}/T_{\rm S}\big]$ will have a negative sign, and its magnitude will gradually increase with decreasing redshifts during these early stages (left panel of Figure \ref{fig:TbPSBS_z}). We do not have a simple analytical model for $\Delta\Tilde{T}_{\rm b}$ in this redshift range, however, it  seems to follow  the global signal as $[\Delta^3]_{\rm Sq}$  is also found to be negative with a slowly evolving magnitude.

{In Model-b IGM heating becomes important at $z \sim 15.5$,  and reduces the magnitude of the absorption signal around the sources by producing warm (or less cold) regions (second panel from left in the middle row of Figure \ref{fig:Tb_map}). Additionally, it also produces small heated regions (where 21-cm signal is in emission) around the same sources. This effect continues as heating progresses and results in an overall reduction in the magnitude of $\Delta\Tilde{T}_{\rm b}$ until $z \sim 15$. The magnitude of the bispectrum also follows this trend for $15.5 \gtrsim z \gtrsim 15$. As heating progresses further by $z \sim 14.5$, the heated regions grow in number and size. The 21-cm signal fluctuations thus can be thought of as being due to a few emission regions embedded in an absorption background. This is the case as predicted in analytical model 2. Hence, the large-scale $\Delta\Tilde{T}_{\rm b}$ will be positive, resulting in a positive $[\Delta^3]_{\rm Sq}$ (i.e. a sign reversal in $[\Delta^3]_{\rm Sq}$). In the redshift range $14.5 \gtrsim z \gtrsim 13$, heated regions grow substantially and percolate. Thus, in this redshift range although the signal fluctuations and its bispectrum cannot be predicted analytically, the simulated $[\Delta^3]_{\rm Sq}$ remains positive by $z \sim 13$.}

As heating progresses the heated regions grow both in number and size and overlap whence they percolate finally resulting in a single large heated region that fills nearly the entire volume. There still remain a few isolated cold absorbing regions embedded in the hot emitting background, and  $\Delta\Tilde{T}_{\rm b}$s thus follows analytical model 5. This results in a negative $[\Delta^3]_{\rm Sq}$, another sign reversal in $[\Delta^3]_{\rm Sq}$, which is indeed seen at $z \sim 12.5$. Due to the continued X-ray heating, the leftover absorption regions disappear (two right-most panels in the second row of Figure \ref{fig:Tb_map}), causing the magnitude of $\Delta\Tilde{T}_{\rm b}$ (and also $[\Delta^3]_{\rm Sq}$) to gradually go down towards the end of the CD.
  
The physical insights obtained from Models-a$_0$, a and b can now be used to understand the evolution of $[\Delta^3]_{\rm Sq}$ for the realistic CD scenario, Model-c. In the redshift range $19 \gtrsim z \gtrsim 14$ $[\Delta^3]_{\rm Sq}$ for Model-c, shown by the dashed green line in the right panel of Figure \ref{fig:TbPSBS_z}, shows similar trends in shape and sign as that of Models-a$_0$ and a. We see that initially Model-c follows Models-a$_0$ and a, implying that Ly$\alpha$ coupling to cold gas is the dominant process during these stages. The sign reversal seen around $z\approx 14$ is due to X-ray heating becoming the dominant process, which remains the case until the end of the simulation at $z=9$, as we see that Model-c follows Model-b. One important point to note here is that during $19 \gtrsim z \gtrsim 11.5$, the magnitude of $[\Delta^3]_{\rm Sq}$ for Model-c is smaller than that of the Models-a$_0$, a and b, simply because the Ly$\alpha$ coupling has not yet saturated in Model-c.

\subsection{Bispectrum a better probe of IGM physics}
We compare the evolution of the signal bispectrum with that of the power spectrum (middle panel of Figure \ref{fig:TbPSBS_z}) in their ability to identify the dominant IGM processes during the CD. We have demonstrated that the bispectrum via its sign and sign changes can conclusively tell us which IGM process dominates $\Delta\Tilde{T}_{\rm b}$ at what cosmic time. The redshift evolution of the power spectrum does show certain subtle features at the time of transition from the dominance of one physical process to the other. However, as this statistic is always positive, it is difficult to unequivocally identify these transitions based on the power spectrum alone.

\subsection{Robustness of the results with \texorpdfstring{$k_1$}{k\_1} modes}
\begin{figure}
\begin{center}
\includegraphics[width=0.8\columnwidth,angle=0]{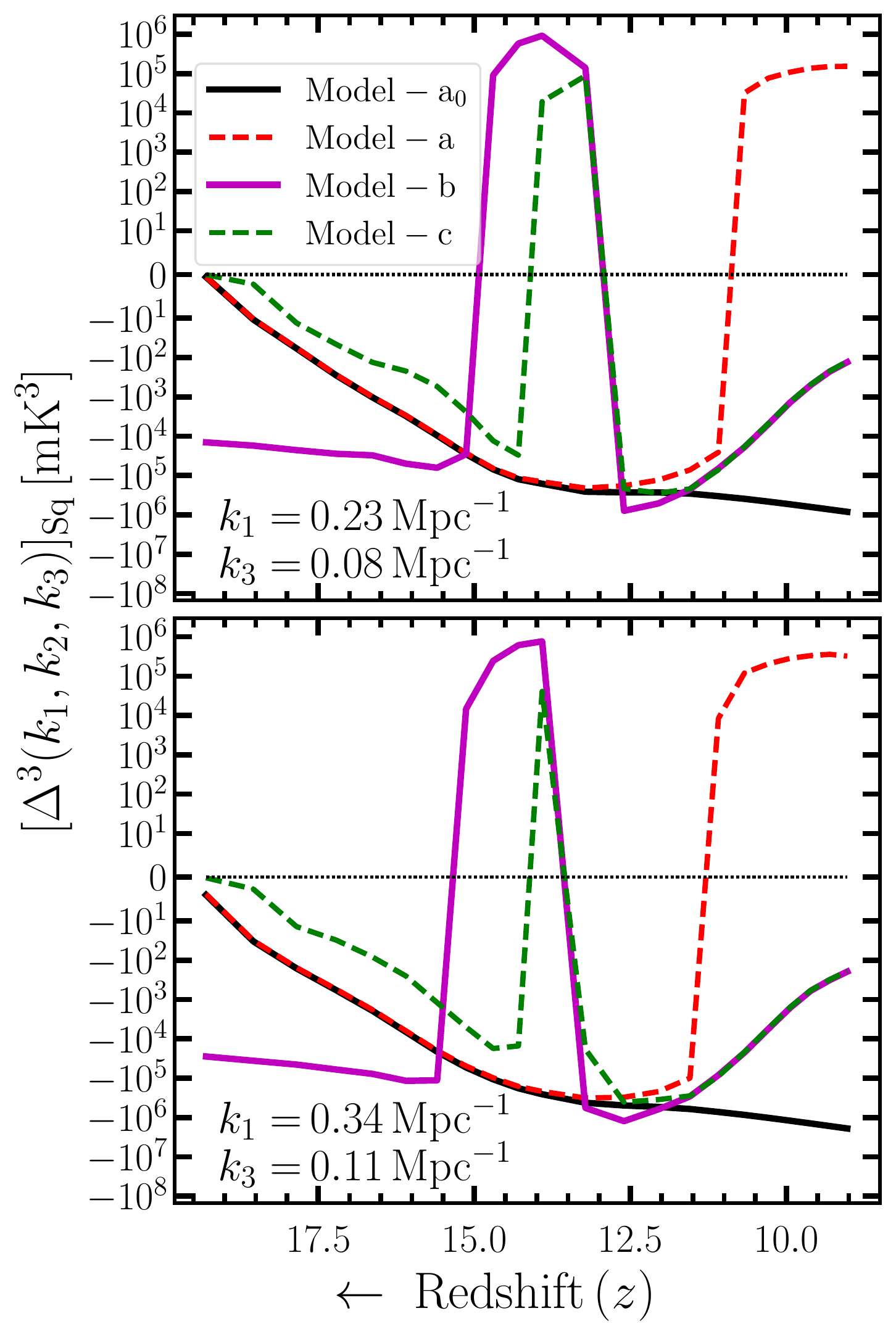}
\caption{Redshift evolution of the bispectrum for squeezed limit $k$-triangle with $k_1 \sim 0.23 \mp$ (top panel) and $k_1 \sim 0.34 \mp$ (bottom panel). Different colours represent different CD scenarios as shown in Table \ref{table1}.}
\label{fig:BS_z_diff_k1}
\end{center}
\end{figure}

We would like to point out that the results that we present here remain the same for a much wider length scale range i.e. for triangles with $k_1 = k_2 \leq 0.34 \mp$ and $k_3 \leq 0.11 \mp$ (see Figure \ref{fig:BS_z_diff_k1}). The robustness of the squeezed limit triangle’s bispectrum with other triangle configurations and with the source parameters during Cosmic Dawn has been thoroughly presented in our recent work \citep{kamran22}.

\subsection{Detectability of the CD 21-cm bispectrum}
Recently, \cite{mondal21} have studied the detectability of the 21-cm bispectrum from the EoR. They have shown that even in the presence of system noise and cosmic variance, a $\geq 5\sigma$ detection of the squeezed limit bispectrum is possible with SKA for $k_1 \lesssim 0.8 \mp$. Since the maximum amplitude of the large-scale squeezed limit 21-cm bispectrum from the CD is $\sim 2$ orders of magnitude larger than that from the EoR {\citep{kamran21}} we, therefore, expect that the large-scale squeezed limit 21-cm bispectrum from the CD would be detectable with the SKA observations. The full detectability prediction is beyond the scope of this letter. We plan to explore it in the future.

\section{Summary and conclusions}
\label{sec:summary}
In this letter, we show for the first time how the two major IGM processes affecting the 21-cm signal during the CD, namely Ly$\alpha$ coupling and X-ray heating, impact the magnitude and sign of large scale ($k_1 \sim 0.16 \mp$) squeezed limit 21-cm bispectrum. We demonstrate that the sign of the $[\Delta^3]_{\rm Sq}$ is negative as long as Ly$\alpha$ coupling is the dominant IGM process. We show that this conclusion holds even in the presence of X-ray heating. The $[\Delta^3]_{\rm Sq}$ goes through a double sign change, around the time when X-ray heating becomes the most dominant IGM process. We have further demonstrated that these transitions cannot be conclusively probed by the power spectrum. 

Our analysis suggests that $[\Delta^3]_{\rm Sq}$ is a better statistic to put constraints on the dominating IGM processes during CD. This will become important in the context of the future 21-cm observations of CD with the SKA. Our analysis here is restricted only to the CD when ionization never becomes the dominant process for the 21-cm fluctuations. In the future, we plan to explore the connection of the sign of $[\Delta^3]_{\rm Sq}$ with the IGM physics during the EoR.

\section*{Acknowledgements}
MK is supported by the foundation Carl Tryggers stiftelse för vetenskaplig forskning, under grant agreement 21:1376 awarded to docent Martin Sahlén. SM and GM are supported by the SPARC Project No. P39, Ministry of Education, Govt. of India and ASEM-DUO India 2020 Fellowship. SM acknowledges financial support through the project titled ``Observing the Cosmic Dawn in Multicolour using Next Generation Telescopes'' funded by the Science and Engineering Research Board (SERB), Department of Science and Technology, Government of India through the Core Research Grant No. CRG/2021/004025. RG is supported by the Israel Science Foundation grant 255/18. GM is supported by Swedish Research Council grant 2020-04691. RM is supported by the Israel Academy of Sciences and Humanities \& Council for Higher Education Excellence Fellowship Program for International Postdoctoral Researchers. ITI is supported by Science and Technology Facilities Council (grants ST/I000976/1 and ST/T000473/1) and Southeast Physics Network (SEP-Net). Some of the numerical results (N-body simulations) were obtained via supercomputing time awarded by PRACE (Partnership for Advanced Computing in Europe) grants 2012061089, 2014102339, 2014102281 and 2015122822 under the project {\sc PRACE4LOFAR}.


\section*{Data Availability}
The simulated data underlying this article will be shared on reasonable request to the corresponding author.





\bibliographystyle{mnras}
\bibliography{reference} 

\begin{thebibliography}{}
\makeatletter
\relax
\def\mn@urlcharsother{\let\do\@makeother \do\$\do\&\do\#\do\^\do\_\do\%\do\~}
\def\mn@doi{\begingroup\mn@urlcharsother \@ifnextchar [ {\mn@doi@}
  {\mn@doi@[]}}
\def\mn@doi@[#1]#2{\def\@tempa{#1}\ifx\@tempa\@empty \href
  {http://dx.doi.org/#2} {doi:#2}\else \href {http://dx.doi.org/#2} {#1}\fi
  \endgroup}
\def\mn@eprint#1#2{\mn@eprint@#1:#2::\@nil}
\def\mn@eprint@arXiv#1{\href {http://arxiv.org/abs/#1} {{\tt arXiv:#1}}}
\def\mn@eprint@dblp#1{\href {http://dblp.uni-trier.de/rec/bibtex/#1.xml}
  {dblp:#1}}
\def\mn@eprint@#1:#2:#3:#4\@nil{\def\@tempa {#1}\def\@tempb {#2}\def\@tempc
  {#3}\ifx \@tempc \@empty \let \@tempc \@tempb \let \@tempb \@tempa \fi \ifx
  \@tempb \@empty \def\@tempb {arXiv}\fi \@ifundefined
  {mn@eprint@\@tempb}{\@tempb:\@tempc}{\expandafter \expandafter \csname
  mn@eprint@\@tempb\endcsname \expandafter{\@tempc}}}

\bibitem[\protect\citeauthoryear{{Barry}, {Wilensky}  \& et al.}{{Barry}
  et~al.}{2019}]{barry19}
{Barry} N.,  {Wilensky} M.,   et al. 2019, \mn@doi [\apj]
  {10.3847/1538-4357/ab40a8}, \href
  {https://ui.adsabs.harvard.edu/abs/2019ApJ...884....1B} {884, 1}

\bibitem[\protect\citeauthoryear{{Behroozi} \& {Silk}}{{Behroozi} \&
  {Silk}}{2015}]{behroozi15}
{Behroozi} P.~S.,  {Silk} J.,  2015, \mn@doi [\apj]
  {10.1088/0004-637X/799/1/32}, \href
  {https://ui.adsabs.harvard.edu/abs/2015ApJ...799...32B} {799, 32}

\bibitem[\protect\citeauthoryear{{Bharadwaj} \& {Pandey}}{{Bharadwaj} \&
  {Pandey}}{2005}]{bharadwaj05a}
{Bharadwaj} S.,  {Pandey} S.~K.,  2005, \mn@doi [\mnras]
  {10.1111/j.1365-2966.2005.08836.x}, \href
  {https://ui.adsabs.harvard.edu/abs/2005MNRAS.358..968B} {358, 968}

\bibitem[\protect\citeauthoryear{{DeBoer}, {Parsons}  \& et al.}{{DeBoer}
  et~al.}{2017}]{deboer17}
{DeBoer} D.~R.,  {Parsons} A.~R.,   et al. 2017, \mn@doi [PAPS]
  {10.1088/1538-3873/129/974/045001}, 129, 045001

\bibitem[\protect\citeauthoryear{{Gallerani}, {Zappacosta}  \& et
  al.}{{Gallerani} et~al.}{2017}]{gallerani17}
{Gallerani} S.,  {Zappacosta} L.,   et al. 2017, \mn@doi [\mnras]
  {10.1093/mnras/stx363}, \href
  {https://ui.adsabs.harvard.edu/abs/2017MNRAS.467.3590G} {467, 3590}

\bibitem[\protect\citeauthoryear{{Ghara}, {Choudhury}  \& {Datta}}{{Ghara}
  et~al.}{2015}]{ghara15a}
{Ghara} R.,  {Choudhury} T.~R.,   {Datta} K.~K.,  2015, \mn@doi [\mnras]
  {10.1093/mnras/stu2512}, \href
  {https://ui.adsabs.harvard.edu/abs/2015MNRAS.447.1806G} {447, 1806}

\bibitem[\protect\citeauthoryear{{Ghara}, {Mellema}, {Giri}, {Choudhury},
  {Datta}  \& {Majumdar}}{{Ghara} et~al.}{2018}]{ghara18}
{Ghara} R.,  {Mellema} G.,  {Giri} S.~K.,  {Choudhury} T.~R.,  {Datta} K.~K.,
  {Majumdar} S.,  2018, \mn@doi [\mnras] {10.1093/mnras/sty314}, \href
  {http://adsabs.harvard.edu/abs/2018MNRAS.476.1741G} {476, 1741}

\bibitem[\protect\citeauthoryear{{Giri}, {D'Aloisio}, {Mellema}, {Komatsu},
  {Ghara}  \& {Majumdar}}{{Giri} et~al.}{2019}]{giri19}
{Giri} S.~K.,  {D'Aloisio} A.,  {Mellema} G.,  {Komatsu} E.,  {Ghara} R.,
  {Majumdar} S.,  2019, \mn@doi [\jcap] {10.1088/1475-7516/2019/02/058}, \href
  {https://ui.adsabs.harvard.edu/abs/2019JCAP...02..058G} {2019, 058}

\bibitem[\protect\citeauthoryear{{Harnois-D{\'e}raps}, {Pen}, {Iliev}, {Merz},
  {Emberson}  \& {Desjacques}}{{Harnois-D{\'e}raps} et~al.}{2013}]{Harnois12}
{Harnois-D{\'e}raps} J.,  {Pen} U.-L.,  {Iliev} I.~T.,  {Merz} H.,  {Emberson}
  J.~D.,   {Desjacques} V.,  2013, \mn@doi [\mnras] {10.1093/mnras/stt1591},
  \href {http://adsabs.harvard.edu/abs/2013MNRAS.436..540H} {436, 540}

\bibitem[\protect\citeauthoryear{{Hinshaw}, {Larson}, {Komatsu}  \& et
  al.}{{Hinshaw} et~al.}{2013}]{hinshaw13}
{Hinshaw} G.,  {Larson} D.,  {Komatsu} E.,   et al. 2013, \mn@doi [\apjs]
  {10.1088/0067-0049/208/2/19}, \href
  {https://ui.adsabs.harvard.edu/abs/2013ApJS..208...19H} {208, 19}

\bibitem[\protect\citeauthoryear{{Hutter}, {Watkinson}, {Seiler}, {Dayal},
  {Sinha}  \& {Croton}}{{Hutter} et~al.}{2020}]{hutter19}
{Hutter} A.,  {Watkinson} C.~A.,  {Seiler} J.,  {Dayal} P.,  {Sinha} M.,
  {Croton} D.~J.,  2020, \mn@doi [\mnras] {10.1093/mnras/stz3139}, \href
  {https://ui.adsabs.harvard.edu/abs/2020MNRAS.492..653H} {492, 653}

\bibitem[\protect\citeauthoryear{{Iliev}, {Mellema}  \& et al.}{{Iliev}
  et~al.}{2008}]{iliev08}
{Iliev} I.~T.,  {Mellema} G.,   et al. 2008, \mn@doi [\mnras]
  {10.1111/j.1365-2966.2007.12629.x}, \href
  {https://ui.adsabs.harvard.edu/abs/2008MNRAS.384..863I} {384, 863}

\bibitem[\protect\citeauthoryear{{Jensen}, {Datta}  \& et al.}{{Jensen}
  et~al.}{2013}]{jensen13}
{Jensen} H.,  {Datta} K.~K.,   et al. 2013, \mn@doi [\mnras]
  {10.1093/mnras/stt1341}, \href
  {http://adsabs.harvard.edu/abs/2013MNRAS.435..460J} {435, 460}

\bibitem[\protect\citeauthoryear{{Kamran}, {Ghara}, {Majumdar}  \& et
  al.}{{Kamran} et~al.}{2021}]{kamran21}
{Kamran} M.,  {Ghara} R.,  {Majumdar} S.,   et al. 2021, \mn@doi [\mnras]
  {10.1093/mnras/stab216}, \href
  {https://ui.adsabs.harvard.edu/abs/2021MNRAS.502.3800K} {502, 3800}

\bibitem[\protect\citeauthoryear{{Kamran}, {Ghara}, {Majumdar}, {Mellema},
  {Bharadwaj}, {Pritchard}, {Mondal}  \& {Iliev}}{{Kamran}
  et~al.}{2022}]{kamran22}
{Kamran} M.,  {Ghara} R.,  {Majumdar} S.,  {Mellema} G.,  {Bharadwaj} S.,
  {Pritchard} J.~R.,  {Mondal} R.,   {Iliev} I.~T.,  2022, \mn@doi [\jcap]
  {10.1088/1475-7516/2022/11/001}, \href
  {https://ui.adsabs.harvard.edu/abs/2022JCAP...11..001K} {2022, 001}

\bibitem[\protect\citeauthoryear{Kolopanis, Jacobs  \& et al.}{Kolopanis
  et~al.}{2019}]{kolopanis19}
Kolopanis M.,  Jacobs D.,   et al. 2019, \mn@doi [\apj]
  {10.3847/1538-4357/ab3e3a}, 883, 133

\bibitem[\protect\citeauthoryear{{Koopmans}, {Pritchard}  \& et al.}{{Koopmans}
  et~al.}{2015}]{koopmans15}
{Koopmans} L.,  {Pritchard} J.,   et al. 2015, in Advancing Astrophysics with
  the Square Kilometre Array (AASKA14). p.~1

\bibitem[\protect\citeauthoryear{{Ma}, {Ciardi}, {Eide}  \& et al.}{{Ma}
  et~al.}{2021}]{ma21}
{Ma} Q.-B.,  {Ciardi} B.,  {Eide} M.~B.,   et al. 2021, \mn@doi [\apj]
  {10.3847/1538-4357/abefd5}, \href
  {https://ui.adsabs.harvard.edu/abs/2021ApJ...912..143M} {912, 143}

\bibitem[\protect\citeauthoryear{{Majumdar}, {Pritchard}, {Mondal},
  {Watkinson}, {Bharadwaj}  \& {Mellema}}{{Majumdar} et~al.}{2018}]{majumdar18}
{Majumdar} S.,  {Pritchard} J.~R.,  {Mondal} R.,  {Watkinson} C.~A.,
  {Bharadwaj} S.,   {Mellema} G.,  2018, \mn@doi [\mnras]
  {10.1093/mnras/sty535}, \href
  {https://ui.adsabs.harvard.edu/abs/2018MNRAS.476.4007M} {476, 4007}

\bibitem[\protect\citeauthoryear{{Majumdar}, {Kamran}, {Pritchard}  \& et
  al.}{{Majumdar} et~al.}{2020}]{majumdar20}
{Majumdar} S.,  {Kamran} M.,  {Pritchard} J.~R.,   et al. 2020, \mn@doi
  [\mnras] {10.1093/mnras/staa3168}, \href
  {https://ui.adsabs.harvard.edu/abs/2020MNRAS.499.5090M} {499, 5090}

\bibitem[\protect\citeauthoryear{{Martocchia}, {Piconcelli}  \& et
  al.}{{Martocchia} et~al.}{2017}]{martocchia17}
{Martocchia} S.,  {Piconcelli} E.,   et al. 2017, \mn@doi [\aap]
  {10.1051/0004-6361/201731314}, \href
  {https://ui.adsabs.harvard.edu/abs/2017A%26A...608A..51M} {608, A51}

\bibitem[\protect\citeauthoryear{Mellema, Iliev, Pen  \& Shapiro}{Mellema
  et~al.}{2006}]{mellema06}
Mellema G.,  Iliev I.~T.,  Pen U.-L.,   Shapiro P.~R.,  2006, \mn@doi [\mnras]
  {10.1111/j.1365-2966.2006.10919.x}, 372, 679

\bibitem[\protect\citeauthoryear{Mertens, Mevius  \& et al.}{Mertens
  et~al.}{2020}]{mertens20}
Mertens F.~G.,  Mevius M.,   et al. 2020, \mn@doi [\mnras]
  {10.1093/mnras/staa327}, \href
  {https://ui.adsabs.harvard.edu/abs/2020MNRAS.493.1662M} {493, 1662}

\bibitem[\protect\citeauthoryear{{Mondal}, {Bharadwaj}  \& et al.}{{Mondal}
  et~al.}{2015}]{mondal15}
{Mondal} R.,  {Bharadwaj} S.,   et al. 2015, \mn@doi [\mnras]
  {10.1093/mnrasl/slv015}, \href
  {http://adsabs.harvard.edu/abs/2015MNRAS.449L..41M} {449, L41}

\bibitem[\protect\citeauthoryear{{Mondal}, {Mellema}, {Shaw}, {Kamran}  \&
  {Majumdar}}{{Mondal} et~al.}{2021}]{mondal21}
{Mondal} R.,  {Mellema} G.,  {Shaw} A.~K.,  {Kamran} M.,   {Majumdar} S.,
  2021, \mn@doi [\mnras] {10.1093/mnras/stab2900}, \href
  {https://ui.adsabs.harvard.edu/abs/2021MNRAS.508.3848M} {508, 3848}

\bibitem[\protect\citeauthoryear{{Paciga}, {Albert}  \& et al.}{{Paciga}
  et~al.}{2013}]{paciga13}
{Paciga} G.,  {Albert} J.~G.,   et al. 2013, \mn@doi [\mnras]
  {10.1093/mnras/stt753}, \href
  {http://adsabs.harvard.edu/abs/2013MNRAS.433..639P} {433, 639}

\bibitem[\protect\citeauthoryear{{Planck Collaboration}, {Ade}, {Aghanim},
  {Armitage-Caplan}, {Arnaud},   \& et al.}{{Planck Collaboration}
  et~al.}{2014}]{planck14}
{Planck Collaboration} {Ade} P.~A.~R.,  {Aghanim} N.,  {Armitage-Caplan} C.,
  {Arnaud} M.,    et al. 2014, \mn@doi [\aap] {10.1051/0004-6361/201321591},
  \href {http://adsabs.harvard.edu/abs/2014A%26A...571A..16P} {571, A16}

\bibitem[\protect\citeauthoryear{{Pritchard} \& {Loeb}}{{Pritchard} \&
  {Loeb}}{2008}]{pritchard08}
{Pritchard} J.~R.,  {Loeb} A.,  2008, \mn@doi [\prd]
  {10.1103/PhysRevD.78.103511}, \href
  {http://adsabs.harvard.edu/abs/2008PhRvD..78j3511P} {78, 103511}

\bibitem[\protect\citeauthoryear{{Pritchard} \& {Loeb}}{{Pritchard} \&
  {Loeb}}{2012}]{pritchard12}
{Pritchard} J.~R.,  {Loeb} A.,  2012, \mn@doi [Rep. Prog. Phys.]
  {10.1088/0034-4885/75/8/086901}, \href
  {https://ui.adsabs.harvard.edu/abs/2012RPPh...75h6901P} {75, 086901}

\bibitem[\protect\citeauthoryear{{Sun} \& {Furlanetto}}{{Sun} \&
  {Furlanetto}}{2016}]{sun16}
{Sun} G.,  {Furlanetto} S.~R.,  2016, \mn@doi [\mnras] {10.1093/mnras/stw980},
  \href {https://ui.adsabs.harvard.edu/abs/2016MNRAS.460..417S} {460, 417}

\bibitem[\protect\citeauthoryear{{Watkinson} \& {Pritchard}}{{Watkinson} \&
  {Pritchard}}{2015}]{watkinson15}
{Watkinson} C.~A.,  {Pritchard} J.~R.,  2015, \mn@doi [\mnras]
  {10.1093/mnras/stv2010}, \href
  {http://adsabs.harvard.edu/abs/2015MNRAS.454.1416W} {454, 1416}

\bibitem[\protect\citeauthoryear{{Watkinson}, {Giri}, {Ross}, {Dixon}, {Iliev},
  {Mellema}  \& {Pritchard}}{{Watkinson} et~al.}{2019}]{watkinson19}
{Watkinson} C.~A.,  {Giri} S.~K.,  {Ross} H.~E.,  {Dixon} K.~L.,  {Iliev}
  I.~T.,  {Mellema} G.,   {Pritchard} J.~R.,  2019, \mn@doi [\mnras]
  {10.1093/mnras/sty2740}, \href
  {https://ui.adsabs.harvard.edu/abs/2019MNRAS.482.2653W} {482, 2653}

\bibitem[\protect\citeauthoryear{{Watkinson}, {Greig}  \&
  {Mesinger}}{{Watkinson} et~al.}{2022}]{watkinson21}
{Watkinson} C.~A.,  {Greig} B.,   {Mesinger} A.,  2022, \mn@doi [\mnras]
  {10.1093/mnras/stab3706}, \href
  {https://ui.adsabs.harvard.edu/abs/2022MNRAS.510.3838W} {510, 3838}

\makeatother
\end{thebibliography}




\label{lastpage}
\end{document}